\documentclass[floatfix,prl,aps,twocolumn,showpacs,preprintnumbers,amsmath,amssymb]{revtex4}

\usepackage{graphicx}
\usepackage{dcolumn}
\usepackage{bm}

\begin{document}
\title{Nonlinear Dynamics of a Bose-Einstein Condensate in a Magnetic Waveguide}
%\title{Nonlinear Dynamics of a Bose-Einstein Condensate in a Magnetic Waveguide}
\author{H. Ott}
\email{ott@pit.physik.uni-tuebingen.de}
\author{J. Fort\'agh}
\author{S. Kraft}
\author{A. G\"unther}
\author{D. Komma}
\author{C. Zimmermann}

\affiliation{Physikalisches Institut der Universit\"at T\"ubingen\\
Auf der Morgenstelle 14, 72076 T\"ubingen, Germany}

\date{\today}
\begin{abstract}
We have studied the internal and external dynamics of a
Bose-Einstein condensate in an anharmonic magnetic waveguide. An
oscillating condensate experiences a strong coupling between the
center of mass motion and the internal collective modes. Due to
the anharmonicity of the magnetic potential, not only the center
of mass motion shows harmonic frequency generation, but also the
internal dynamics exhibit nonlinear frequency mixing. Thereby, the
condensate shows shape oscillations with an extremely large change
in the aspect ratio of up to a factor of 10. We describe the data
with a theoretical model to high accuracy. For strong excitations
we test the experimental data for indications of a chaotic
behavior.

\end{abstract}

\maketitle

Since the first  realization  of Bose-Einstein condensates in 1995
ideas have been discussed how to exploit this new form of coherent
matter for atom optics.  Atom interferometry with single atoms has
been extensively investigated in the past decade
\cite{Berman1997a} and finds important applications for high
precision measurements of gravitational or rotational forces
\cite{Riehle1991a,Kasevich1991a}. It was also possible to realize
and demonstrate a number of atom optical elements such as mirrors
\cite{Johnson1998a}, gratings \cite{Rasel1995a}, beamsplitters
\cite{Cassettari2000a} and waveguides \cite{Key2000a,Muller1999a}.
It is a fascinating vision to integrate these elements on a
microfabricated atom chip. Extremely sensitive chip based sensors
for forces and accelerations as well as for applications in
quantum computing are conceivable. Since Bose-Einstein condensates
have been loaded into microfabricated traps
\cite{Ott2001a,Hansel2001b} this vision has changed into a very
real research topic. Integrated atom optics with condensates
combine the nonlinear interaction of a Bose-Einstein condensates
with trapping potentials, whose geometry is in general anharmonic.
This is in contrast to both single atom experiments, where the
interaction plays no important role, and conventional experimental
arrangements for condensates, where the trapping potential is
typically harmonic and consequently the center of mass (CM) motion
is decoupled from the internal dynamics \cite{Dobson1994a}. In
microtraps one therefore expects not only a coupling between the
external and internal dynamics but also pronounced nonlinear
effects including even chaotic dynamics for large excitation
amplitudes. The study of the dynamics of a condensate in
micropotentials is an extension of previous work on collective
excitations and provides essential information for future
applications for integrated atom optics.

We present a detailed experimental study of the condensate
dynamics in the most fundamental atom optical element --- a
magnetic waveguide. We regard waveguides which are closed at both
ends such that the generic motion of the condensate is a periodic
oscillation. The linear propagation of Bose-Einstein condensates
along a non terminated guide was recently investigated in
Refs.\,\cite{Leanhardt2002a} and \cite{Fortagh2002c}. We have
measured the time evolution of the CM motion and of the aspect
ratio i.\,e. the ratio between the radial and the axial condensate
radii. In contrast to previous experimental work on collective
modes of a condensate, we realize a closed system with conserved
energy and a self-sustained excitation scheme. This system allows
us to realize high amplitude oscillations of the aspect ratio
which exceed previously reported values by far. To describe the
experimental data we have developed a theoretical model which is
described in Ref.\,\cite{Ott2002b}. The model is based on the
separation of the CM motion and reduces the internal dynamics to
that of a condensate in a fictitious time-dependent harmonic
potential. Such a scenario is known to be exactly solvable in
Thomas-Fermi approximation \cite{Kagan1997b,Castin1996a}. The
model also allows us to identify an experimentally accessible
regime where the dynamics become chaotic.

\begin{figure}
\begin{center}
\includegraphics[width=7cm]{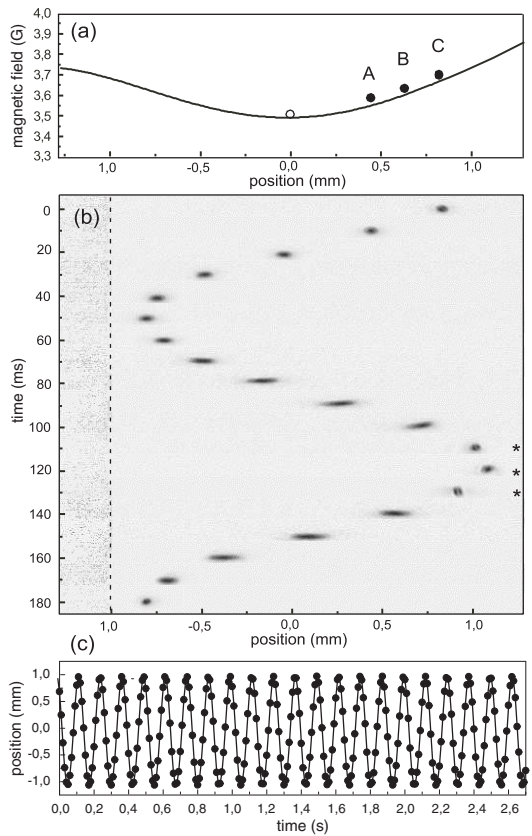}
\caption{\label{figure1}Oscillation of the condensate in an
anharmonic waveguide. The condensate is trapped below the
microstructure which generates the magnetic field for the trapping
potential. (a) Magnetic field in the trap minimum along the axial
direction. The black dots indicate the starting positions of the
oscillation for each experimental series. The circle marks the
initial position without displacement. (b) Absorption images of
the condensate after 20\,ms time of flight (series C, 10\,ms
intervals). Due to the time of flight, the amplitude of the
oscillation is enhanced with respect to oscillation in the trap
(also a phase shift occurs). For condensates marked with an
asterisk the absorption images show diffraction and the aspect
ratio can not be determined properly. (c) CM motion of the
condensate (series B): experimental data (dots) and fit with three
sinusoidal functions (line).}
\end{center}
\end{figure}

The waveguide potential is generated with microfabricated copper
conductors at the surface of a chip. The chip is mounted upside
down in a vacuum chamber at a background pressure of
$10^{-11}$\,mbar. The $^{87}$Rb atoms are collected in a six-beam
magneto-optical trap, pumped into the $(F=2,m_{\rm{F}}=2)$ ground
state and trapped in a spherical magnetic quadrupole trap. The
atoms are then adiabatically transferred into the waveguide
potential at the chip surface by gradually changing the magnetic
field geometry \cite{Ott2001a,Fortagh2002b}. By forced
evaporation, Bose-Einstein condensates with 100.000 atoms are
produced. For the presented experiments, the trapping frequencies
are adjusted to $110$\,Hz in the radial and $8$\,Hz in the axial
direction. The shape of the axial potential is sketched in
Fig.\,\ref{figure1}a. The waveguide potential is characterized by
an almost constant radial oscillation frequency (the maximal
deviations are smaller than 10 percent) and a strongly varying
axial curvature. With a numerical calculation of the magnetic
field, we determine the axial potential $U(x)$ to
\begin{equation}\label{eq5}
\frac{U(x)}{\mu}\left[G\right]=0.219x^2-0.013x^3-0.033x^4-0.007x^5,
\end{equation}
where $x$ has to be quoted in millimeter and $\mu$ is the magnetic
moment of the atom. To induce the oscillation we displace the
condensate within $400$\,ms with a magnetic field gradient and
switch off the field gradient non-adiabatically
($<60\,\mu\mathrm{s}$). The experiment was performed for three
different initial displacements $d$: $0.45$\,mm (series A,
duration $t_{\rm A}=1.015$\,s), $0.63$\,mm (B, $t_{\rm B}=2.7$\,s)
and $0.79$\,mm (C, $t_{\rm C}=0.765$\,s). In Fig.\,\ref{figure1}b
the first $1.5$ oscillation periods of series C are shown. To
determine the CM position and the aspect ratio, the absorption
images are fitted to a Thomas-Fermi profile. The aspect ratio
bears the advantage of being independent from the atom number. To
exclude influences of the nearby copper conductors
\cite{Fortagh2002d,Kraft2002a}, the experiment was carried out in
a distance to the surface of $220$\,$\mu\rm{m}$.

\begin{figure}
\centering
\includegraphics[width=8cm]{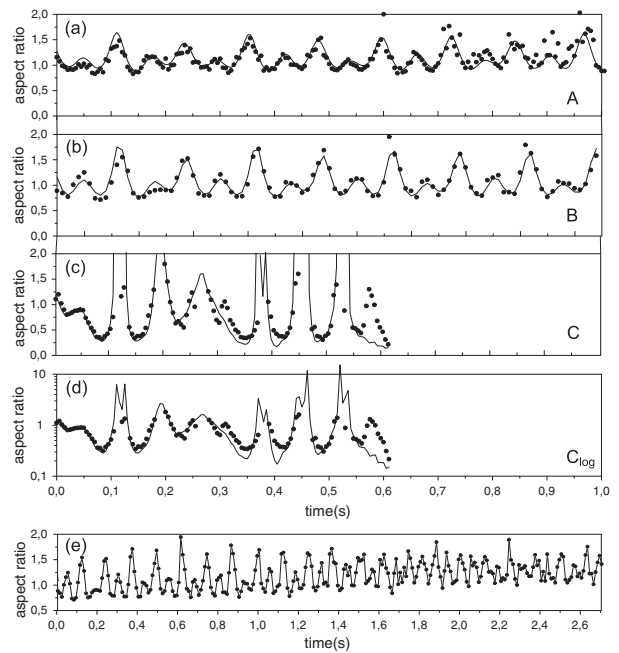}
\caption{\label{figure2}Evolution of the aspect ratio of the
condensate for the data series A (a), B (b), and C (c) (dots) and
theoretical model (solid line). (d) Plot (c) with a logarithmic
scale. (e) All data points of series B.}
\end{figure}

Our theoretical model describes the condensate with a set of two
coupled differential equation systems. The first system are the
classical equations of motion for the CM position of the
condensate
\begin{equation}\label{eq1}
m\ddot{\mathbf{R}}=-\nabla U(\mathbf{R}),
\end{equation}
where $m$ is the mass of the atom, $\mathbf{R}$ is the CM position
and $U$ is the confining potential. In Thomas-Fermi approximation,
the density distribution of a condensate in a time-dependent
harmonic potential can be written as follows \cite{Castin1996a}:
\begin{equation}
n(\mathbf{x,t})=\frac{\mu}{g}\frac{1}{\lambda_1(t)\lambda_2(t)\lambda_3(t)}\left(1-\sum_{i=1}^3\left(\frac{x_i}{r_{i0}\lambda_i(t)}\right)^2\right),
\end{equation}
where $\mu$ is the chemical potential and $g$ is the coupling
constant. The $\lambda_i(t)$ contain the complete time evolution
of the Thomas-Fermi radii $r_i(t)=r_{i0}\lambda_i(t)$. The
$\lambda_i$ are governed by three ordinary differential equations
\begin{equation}\label{eq2}
\ddot{\lambda}_i=\frac{\omega_{i0}^2}{\lambda_i\lambda_1\lambda_2\lambda_3}-\omega_i^2(t)\lambda_i,\,\,\,\,\,\,\,i=1,2,3.
\end{equation}
Here, $\omega_i(t)$ are the time-dependent oscillation frequencies
of the harmonic potential. If the modulation of the potential
starts at $t=0$, the initial conditions for the parameters are
$\omega_i(0)=\omega_{i0}$ and $r_i(0)=r_{i0}$ with
$\lambda_i(0)=1$. The anharmonicity of the potential now provides
a coupling of the CM motion and the time evolution of the
Thomas-Fermi radii, which is given by
\begin{equation}\label{coupling}
\omega_i^2(t)=\frac{1}{m}\frac{\partial^2}{\partial
x_i^2}U(\mathbf{R}),\,\,\,\,\,\,\,i=1,2,3.
\end{equation}
To compare the experimental data with the model, (\ref{eq1}),
(\ref{eq2}) and (\ref{coupling}) are numerically solved with the
initial conditions $\lambda_i(0)=1$ and $\mathbf{R}(0)=(d,0,0)$.

The experimental data for the CM motion of series B are presented
in Fig.\,\ref{figure1}c . We cannot extract any damping and the
data are consistent with a ($1/e$)\,--\,lifetime of at least 20
minutes. Due to the anharmonicity of the potential (\ref{eq5}),
the frequency spectrum of the CM motion shows second and third
harmonic generation. For series C with the largest initial
displacement, the nonlinear effects are most pronounced and lead
to a relative amplitude of 5 percent for the second harmonic and 1
percent for the third harmonic frequency with respect to the
amplitude for the fundamental frequency $\nu_0$. For small
amplitudes $\nu_0$ amounts to $8.5$\,Hz but is reduced with
increasing displacement (A, B, and C) by 4, 6, and 11 percent
respectively. Because of the extremely low damping, the CM motion
is suited to detect small forces, which may shift the trapping
potential. An estimate for our experimental parameters indicates a
resulting sensitivity of $F/m=10^{-4}g$, with the gravitational
acceleration $g$.

In Fig.\,\ref{figure2}a-\ref{figure2}d the measured aspect ratio
is compared with the theoretical model. While the theory contains
no free parameter, a constant multiplication factor of $1.2$ is
needed to describe the data accurately. This factor arises from
the fact, that the condensate is radially compressed during the
switch-off of the magnetic trap which --- due to the mean field
repulsion --- leads to a larger radial size in the time-of-flight
image \cite{switchoff}. The agreement between experiment and
theory is remarkable and proves true the picture that the external
motion in an anharmonic potential leads to a time-dependent
harmonic potential in the rest frame. In Fig.\,\ref{figure1}b, the
condensates marked with an asterisk cannot be spatially resolved
by our absorption imaging system, indicating an axial extension of
the condensate of less than $15\,\mu\mathrm{m}$. Therefore, the
condensate may in fact follow the strong dynamics, which is
predicted by the theory (Fig\,\ref{figure2}d). For the present
data, we measure a change in the aspect ratio by more than a
factor of 10, the largest value reported so far. The data of
series B which cover $2.7$\,s are promising to find effects
arising from loss of atoms (Fig\,\ref{figure2}e). At the end of
the series, the atom number has dropped by more than a factor of
15 with only 3000 atoms left in the condensate. The systematic
increase of the aspect ratio is due to the crossover to the 1D
regime as defined in Ref.\,\cite{Goerlitz2001a}. The Thomas-Fermi
approximation breaks down because the chemical potential of the
condensate becomes smaller than the radial kinetic energy, which
then dominates the radial expansion of the condensate.

\begin{figure}
\centering
\includegraphics[width=8cm]{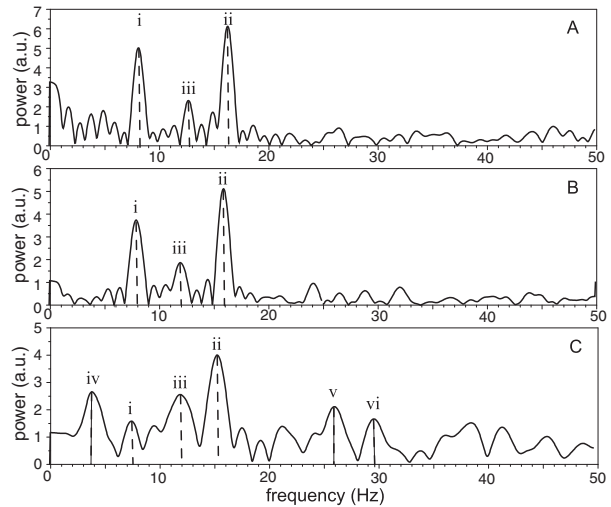}
\caption{\label{figure3}Frequency spectra of the aspect ratio. The
frequencies of the labeled peaks are: (i) fundamental frequency
$\nu_0$, (ii) second harmonic frequency $2\nu_0$, (iii) quadrupole
resonance frequency $\nu_q\simeq\sqrt{5/2}\,\nu_0$, (iv)
difference frequency $2\nu_0-\nu_q$, (v) sum frequency
$2\nu_0+\nu_q$ and (vi) fourth harmonic frequency $4\nu_0$.}
\end{figure}

We have performed a frequency analysis of the aspect ratio with an
algorithm suitable for unequally spaced or missing data
\cite{Lomb1976a}. The results are shown in Fig.\,\ref{figure3}.
The spectra show three major frequencies: the fundamental
frequency $\nu_0$ (peak (i) in Fig.\,\ref{figure3}), the second
harmonic (ii) and the resonance frequency $\nu_q$ of the lowest
lying collective mode (iii). The latter does not deviate more than
4 percent from its small amplitude value for strongly anisotropic
traps which is given by $\sqrt{5/2}\,\nu_0$. Furthermore, the
spectrum for series C (Fig.\,\ref{figure3}) shows several mixed
frequencies. Two of them, the sum and the difference between the
second harmonic frequency and the lowest resonance frequency
$2\nu_0-\nu_q$ and $2\nu_0+\nu_q$ have been indentified ((iv) and
(v)). This gives evidence for the nonlinear coupling of the
collective excitations. Other than the coupled modes observed by
Hechenblaikner {\it et al.} \cite{Hechenblaikner2000a}, the mode
coupling in our experiment is non-resonant, and the mixed
frequencies vanish in the limit of small amplitudes. Because the
condensate performs a free oscillation in an external potential
the total energy of the system is conserved. Summing up only the
contributions for the internal energy of the condensate, one finds
that the sum is not constant. Adding the kinetic and potential
energy of the CM motion, the energy is conserved, which can be
verified by using a Hamiltonian formalism \cite{Ott2002b}.
Therefore, not only the internal dynamics is affected by the
external motion, but there is also an energy exchange between the
internal dynamics and the CM motion. In our experiment this effect
leads to a theoretically calculated amplitude variation for the CM
motion of $1/4000$, which is too small to be detected.

\begin{figure}
\begin{center}
\includegraphics[width=6cm]{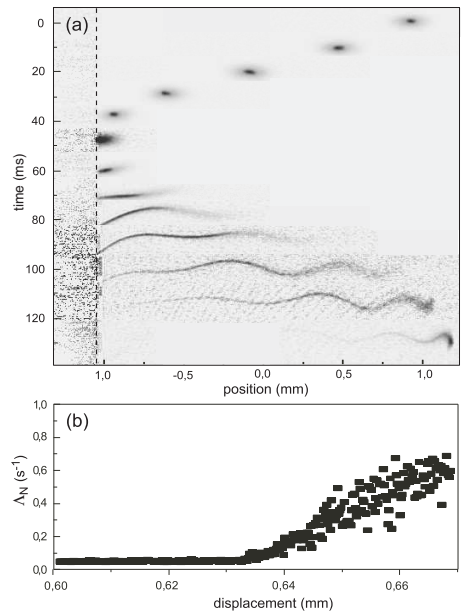}
\caption{\label{figure4}(a) Oscillation for an initial
displacement of $0.94$\,mm. The absorption images are taken after
20\,ms time of flight. The vertical dotted line indicates the
surface of a wire which narrows the display window. (b) Lyapunov
exponent: $\Lambda_{\rm N}=1/(N\tau)\sum_{i=1}^N\ln{d_i/d_0}$ as a
function of the initial displacement $d$. $d_0$ is the initial
Euclidian distance between two neighboring trajectories, $\tau$ is
the duration of one iteration and $N$ is the number of iterations.
The Lyapunov exponent is defined in the limit
$\Lambda=\lim_{N\rightarrow\infty}\Lambda_{\rm N}$. Each pair of
trajectories was numerically solved for $200$\,s ($d_0=0.1\,\mu$m,
$\tau=1\,$ms, $N=2\times10^5$). Above a displacement of $0.67$\,
mm the numerics are no longer stable for the total propagation
time.}
\end{center}
\end{figure}

Fig.\,\ref{figure4}a shows absorption images for an initial
displacement of $0.94$\,mm. The data show a strong axial
excitation of the condensate, which raises the question of the
onset of chaos. A signature of a chaotic behavior in experimental
data is an enhanced noise in the frequency spectrum of the data.
The spectrum of the aspect ratio for series C in
Fig.\,\ref{figure3} clearly shows an enhanced background, compared
to the two other series. However, this should not be regarded as
an unchallengeable experimental proof. Such a proof would require
a longer data set and an enhanced resolution of the imaging
system. Nevertheless, the theoretical model can be analyzed for
the experimental parameters. One of the strongest evidences for a
chaotic dynamic is a positive Lyapunov exponent, which indicates
an exponential sensitivity to the initial conditions. For this
purpose neighboring trajectories for slightly different initial
displacements have been simulated over 200\,s. Fig.\,\ref{figure4}
shows the evolution of the Lyapunov exponent with increasing
displacement. For a displacement of more than $0.63$\,mm, the
calculated Lyapunov exponent becomes positive and grows linearly.
As the total energy of the system is conserved, one can easily
test the numerical stability and exclude any numerical artefacts.
The corresponding Poincar\'e maps and the frequency spectra of the
trajectories indicate the same results. This gives strong
evidence, that series C was performed in a regime where the
longtime behavior of the trajectory is chaotic.

In conclusion, we have studied the dynamics of a Bose-Einstein
condensate oscillating in an anharmonic trapping potential. We
have identified several nonlinear effects including second and
third harmonic generation of the center of mass motion, a coupling
of the internal and external dynamics and nonlinear mode mixing.
The theoretical model describes these effects with high accuracy.
The model predicts a chaotic dynamic above a critical oscillation
amplitude and the experimental data give indication for a chaotic
motion too. Considering applications of magnetic microtraps and
integrated atom optics, we have shown that the motion of a
condensate in a magnetic waveguide necessarily leads to
excitations of the inner degrees of freedom. A thorough
understanding of these effects plays a key role to exploit the
nonlinearity to build high precision sensors. It also shows
fundamental limits for the use of Bose-Einstein condensates for
technical applications.

We gratefully acknowledge financial support from the Deutsche
Forschungsgemeinschaft under Grant No. Zi/419-5.


\begin{references}
\bibitem{switchoff}
Assuming a stepwise change of the radial trapping frequency we can
include the distortion in the theory to model the effect of the
compression. We find e.\,g., that an increase of the radial
trapping frequency to $500$\,Hz for a duration of
$70\,\mu\mathrm{s}$ is sufficient to cause the factor of $1.2$.
\end{references}
\end{document}